# The local structure and optical absorption characteristic investigation on Fe doped $TiO_2$ nanoparticles

Tianxing Zhao(赵天行)[1], Yajuan Feng(冯亚娟)[1], Junheng Huang(黄均衡)[1], Jinfu He(何劲夫)[1], Qinghua Liu(刘庆华)[1], Zhiyun Pan(潘志云)[1*] and Ziyu Wu(吴自玉)[1,2*]

[1]National Synchrotron Radiation Laboratory, University of Science and Technology of China, 42# HeZuoHua South Road, 230029, Hefei, Anhui, P.R. China
[2]Institute of High Energy Physics, Chinese Academy of Sciences, 100049, Beijing, P.R. China
*Corresponding author. *E-mail address: *wuzy@ustc.edu.cn and zhypan@ustc.edu.cn*

**Abstract**  The local structures and optical absorption characteristic of Fe doped $TiO_2$ nanoparticles synthesized by the sol-gel method were characterized by X-ray Diffraction (XRD), X-ray absorption fine structure spectroscopy (XAFS) and UV-Vis absorption spectroscopy (UV-Vis). XRD patterns show that all Fe-doped $TiO_2$ samples have the characteristic anatase structure. Accurate Fe and Ti K-edge EXAFS analysis further reveal that all Fe atoms replace Ti atoms in the anatase lattice. The analysis of UV-Vis data shows a red shift to the visible range. According to the above results, we claim that substitutional Fe atoms lead to the formation of structural defects and new intermediate energy levels appear, narrowing the band gap and extending the optical absorption edge towards the visible region.

## 1 Introduction

In recent years, due to its low cost, non-toxicity, easy-synthesis, long-term stability and high efficiency[1], a lot of attention has been devoted to $TiO_2$ photocatalysts to find solutions useful to treat environmental pollution problems and the continuous increasing demand of energy[2]. However, its narrow range of response in the ultraviolet actually limits large-scale industrial applications[3].

Recently, many researches were carried out to investigate metal ion-doping $TiO_2$ systems to improve the photocatalytic activity.[4] Different studies have reported doping with suitable transitional metals[5-10]. Among them, iron was considered to be one of the most appropriate candidates because the radius of $Fe^{3+}$ (0.645 Å) is similar to the $Ti^{4+}$ (0.605 Å) and $Fe^{3+}$ can be easily incorporated into the crystal lattice of $TiO_2$ [11].

The local structure around iron strongly affects the photocatalytic activity properties of Fe-doped $TiO_2$. As an example, if Fe ions in doped samples are aggregated to $\alpha$-$Fe_2O_3$ precipitates, doping may affect the catalytic effect of $TiO_2$[12]. Therefore, an accurate determination of local structure information around iron is mandatory. X-ray absorption fine structure (XAFS) is a powerful method to investigate the local structure around a specific component and can be used to obtain the local structure information around a selected atom[13].

In this work, the local structure around Fe atoms in doped $TiO_2$ nanoparticles prepared by sol-gel method with different Fe concentrations, were investigated by using XRD and XAFS techniques to identify where Fe ions go inside the $TiO_2$ matrix and to improve the optical absorption.

## 2 Experimental Section

### 2.1 Sample Preparation

Pure and Fe-doped $TiO_2$ nanoparticles ($Fe_xTi_{1-x}O_2$) were prepared by the sol-gel method and the molar ratios of the Fe dopants (x) were 0.6% and 6.0%[14]. Pure amounts of $Fe_2(SO_4)_3$ and $Fe(NO_3)_3$ with a certain mass ratio were dissolved in distilled water under continuous stirring for 2 h and then 0.05 mol/L $Na_2CO_3$ solution was dropped, wisely added to the above solution under continuous stirring for 2 h, and the resulting solution was evaporated by heating and stirring at 353 K for 4 h. The mixture was put into the oven at 393 K for 10 h and then calcined in the muffle at 873 K for 10 h.

### 2.2 Measurement

The UV-Vis absorption spectra were performed with an Analyst 800 in the spectral range from 200 nm to 800 nm. XRD patterns were collected using Cu K$\alpha$ ($\lambda$ = 0.154 nm) radiation in the 2$\theta$ range from 20º to 80°. Fe and Ti K-edge spectrosa of the prepared samples were collected at the U7C beamline of the National Synchrotron Radiation Laboratory (NSRL, China). XAFS data were analyzed by the UWXAFS3.0 software package according to standard data analysis procedures[15].

## 3 Results & Discussion

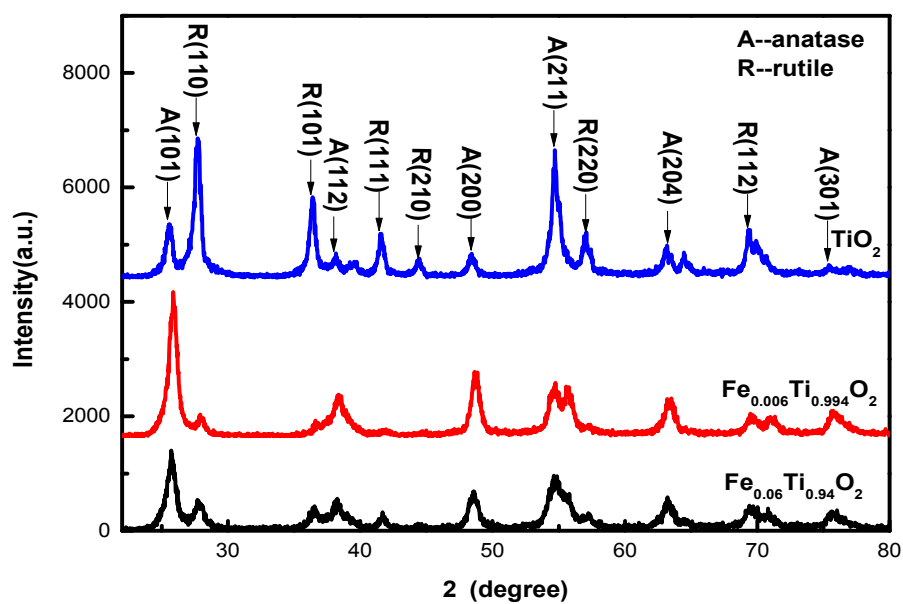

Figure 1. XRD patterns of $Fe_xTi_{1-x}O_2$ nanoparticles (x=0%, 0.6% and 6%)

To detect the crystalline order and/or to identify different ordered phases we performed XRD measurements in these $TiO_2$ nanoparticles doped with different Fe concentrations. From Figure 1, we can see that all samples are characterized by a mixture of an anatase and a rutile phase without diffraction peaks eventually associated to other iron oxide or impurity phases. It is worth mentioning that, compared with the pure sample, the intensity of some peaks of the rutile phase decreases in the doped samples, such as peaks at 2θ= 36.48º, 41.73º, 44.30º, 56.89º, and 69.48º. Data point out that Fe doping could affect the crystallization of the $TiO_2$, especially that of the rutile phase. Regarding the (101) crystalline reflection (2θ=25.3º) of the anatase phase, increasing the content of Fe ions from 0.6% to 6% its intensity increases and shifts to lower angles. This result clearly points out that Fe ions may increase the disorder and expand the lattice size of the structure of $Fe_xTi_{1-x}O_2$ nanoparticles.

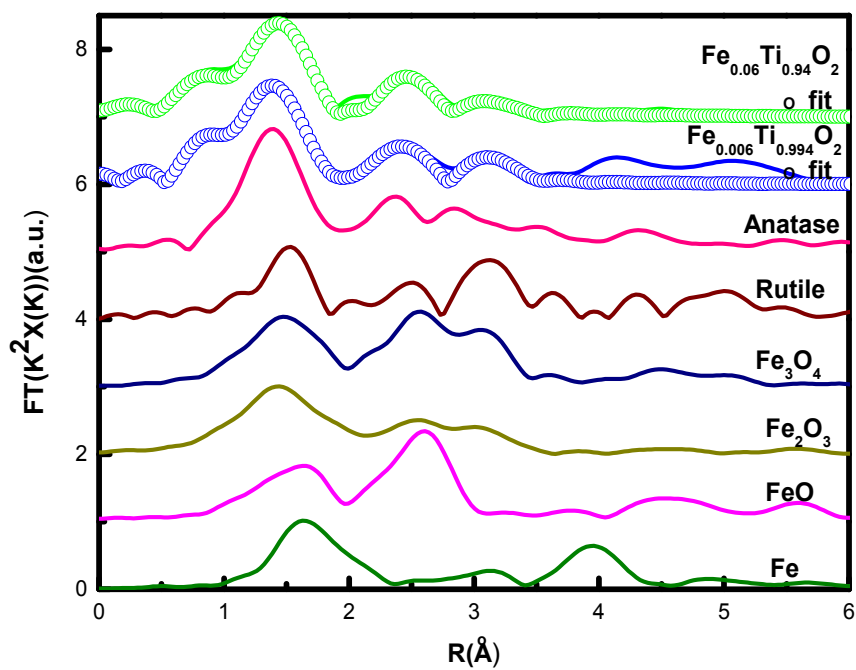

Figure 2. The radial structural functions obtained by the FTs of Fe K-edge EXAFS spectra and their fits, for $Fe_{0.006}Ti_{0.994}O_2$ and $Fe_{0.06}Ti_{0.94}O_2$, and the radial structural functions of the Ti K-edge EXAFS spectra of anatase (theoretical) and rutile (theoretical), as well as Fe K-edge FTs EXAFS of Fe, FeO, $Fe_2O_3$ and $Fe_3O_4$.

Figure 2 displays Ti K-edge EXAFS Fourier transforms (FTs) spectra of anatase, and Fe K-edge EXAFS Fourier transforms (FTs) spectra of $Fe_{0.006}Ti_{0.994}O_2$ and $Fe_{0.06}Ti_{0.94}O_2$ and their fits. As references, the Fe K-edge spectra of Fe, FeO, $Fe_2O_3$, $Fe_3O_4$, and the Ti K-edge EXAFS spectrum of rutile are also showed in Figure 2. The FT features of $Fe_{0.006}Ti_{0.994}O_2$ and $Fe_{0.06}Ti_{0.94}O_2$ nanoparticles are very close to anatase $TiO_2$ powders, with the first coordination peaks of both Fe K-edge and Ti K-edge at ~1.4 Å, a value corresponding to Fe-O and Ti-O coordination shells. The second single coordination peaks at 2.3 Å corresponds to the Fe-O-Ti and the Ti-O-Fe coordination shells. These peaks are clearly different from those of Fe, FeO, $Fe_2O_3$, and $Fe_3O_4$ systems. Therefore, we may rule out the presence of Fe, FeO, $Fe_2O_3$, and $Fe_3O_4$. This also indicates that Fe ions in Fe-doped $TiO_2$ nanoparticles replace Ti atoms in the $TiO_2$ lattice within the doping concentration from 0.006 to 0.06. It is also important to underline that, compared to anatase, the first peak in the $Fe_xTi_{1-x}O_2$ samples presents a shift of ~ 0.1 Å toward high R. This result points out an expansion of the bond length of the Fe-O bond respect to that of Ti-O. In addition, we can see that the weaker intensity of the second coordination peaks suggests that these samples have a specific Fe-O-Ti structure containing many structural defects, which are extremely important to enhance the photocatalytic activity of these nanoparticles[16,17].

In order to obtain additional quantitative structural information, we fit the main peak corresponding to Fe-O and Ti-O pairs. Data obtained using the basic EXAFS formula are shown in Figure 2[18]. The best fit of the structural parameters obtained for these samples are summarized in the Table 1 and the results of the fit are shown as empty circles in Figure 2. As shown in Table 1, we can see that both Fe-O bond lengths in the $Ti_{0.994}Fe_{0.006}O_2$ are ~0.02 Å longer than the Ti-O bond lengths of the anatase $TiO_2$ phase, pointing out that Fe ions replace Ti ions and expand the cluster of of the nearest oxygen atoms[9]. With a doping content increasing from 0.6% to 6%, both Fe-O bond lengths expand, in agreement with XRD data. Moreover, because Fe doping enhances the distortion this phase is thermodynamically less stable phase[19].

**Table 1.** Structural parameters around Fe atoms in the synthesized titania samples.

| Sample | Coordination | Coordination numbers (N) | R(Å) | $\sigma^2(Å^2)$ |
|---|---|---|---|---|
| Theoretical | Ti-O | 4 | 1.92 | |
| $TiO_2$ anatase | Ti-O | 2 | 1.97 | |
| $Ti_{0.994}Fe_{0.006}O_2$ | Fe-O | 3.9±0.1 | 1.94±0.01 | 0.0065±0.0001 |
| | Fe-O | 1.9±0.1 | 1.99±0.01 | 0.0064±0.0002 |
| $Ti_{0.94}Fe_{0.06}O_2$ | Fe-O | 3.9±0.1 | 1.96±0.01 | 0.0073±0.0003 |
| | Fe-O | 1.8±0.1 | 2.03±0.01 | 0.0073±0.0002 |

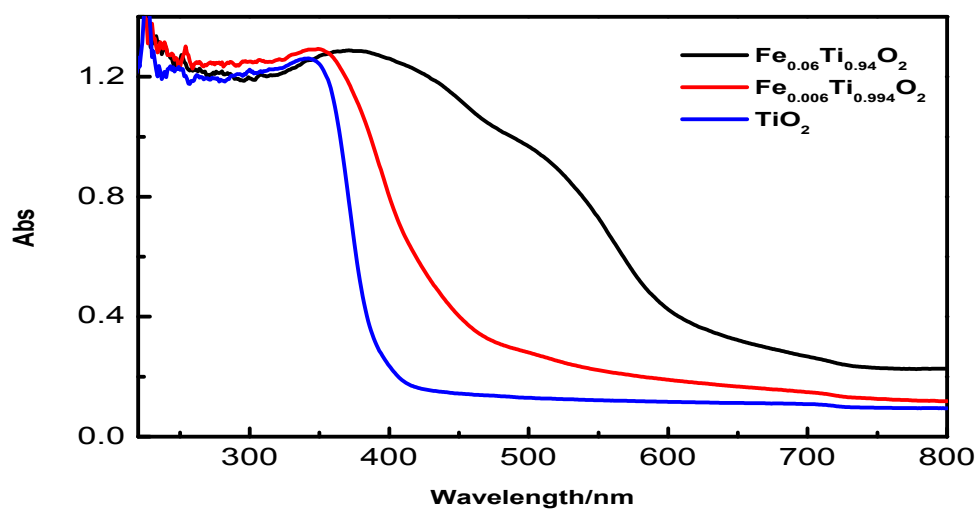

Figure 3. UV-Vis absorption spectra of $Fe_xTi_{1-x}O_2$ nanoparticles for x=0, 0.6 and 6%.

To investigate the optical absorption characteristic of Fe-doped $TiO_2$ nanoparticles, UV-Vis absorption spectra were also collected. As shown in Figure 3, the light absorption edges of 0.6% and 6% Fe-doped $TiO_2$ are 490 nm and 613 nm, respectively, with a remarkable red shift to the visible range compared to the spectrum of pure $TiO_2$ (387 nm). At the same time, the absorption intensity of visible radiation increases when the iron doping content increased from 0.6% to 6%.

With the increase of Fe ions, the absorbance in the visible region is enhanced, a behavior associated to the Fe ion occupancy in the Ti sites of the $TiO_2$ lattice. Thus an interaction among d electrons of Fe and the $TiO_2$ conduction or valence band occurs[20] eventually narrowing the energy gap of the titanium oxide through the formation of new intermediate energy levels[21,22].

## 4 Conclusion

The local structure of Fe doped $TiO_2$ nanoparticles prepared by a sol-gel method have been investigated by XRD and XAFS, combined with UV-Vis to monitor their absorption characteristic. XRD patterns show that all Fe-doped $TiO_2$ samples qualitatively have the anatase $TiO_2$ structure. Moreover, detailed Fe and Ti K-edge EXAFS experiments further reveal that Fe atoms replace Ti atoms in the anatase lattice. Doping with iron leads to the formation of the structural defects and, probably, a lot of intermediate energy levels that may narrow the energy gap being, the mechanism responsible of the red shift observed in the spectra of these materials in the UV-Vis region.

## 5 Acknowledgments

This work was supported by the National Basic Research Program of China (No. 2012CB825801), the Science Fund for Creative Research Groups of the NSFC (11321503), the National Natural Science Foundation of China (No. 11321503, 11179004) and the Guangdong Natural Science Foundation (No. S2011040003985). Authors would like to thank the NSRL facilities for beamtime assignment.